\documentclass[a4paper]{jpconf}
\usepackage{graphicx}

\usepackage{hyperref}
\usepackage{amssymb}
\usepackage{longtable}
\usepackage{rotating}
\usepackage{color}
\usepackage{epsfig}
\usepackage{epsf}
\usepackage{fancyhdr}

\def\be{\begin{equation}}
\def\ee{\end{equation}}
\def\th{\textrm{\mbox{\tiny{th}}}}

\def\Tcoh{T_{\textrm{\mbox{\tiny{coh}}}}}

\def\bea{\begin{eqnarray}}
\def\eea{\end{eqnarray}}
\def\N{\textit{\mbox{\tiny{N}}}}

\begin{document}

\title[Improved Hough search for gravitational wave pulsars]
     {Improved Hough search for gravitational wave pulsars}

\author{Alicia M. Sintes\dag\ddag, and Badri Krishnan\ddag}

\address{\dag\ Departament de F\'{\i}sica, Universitat de les Illes
Balears, Cra. Valldemossa Km. 7.5, E-07122 Palma de Mallorca,
Spain}

\address{\ddag\ Max-Planck-Institut f\"ur
    Gravitationsphysik, Albert Einstein Institut, Am M\"uhlenberg 1,
    D-14476 Golm, Germany}

\ead{sintes@aei.mpg.de,  badri.krishnan@aei.mpg.de }

\begin{abstract}
  We describe an improved version of the Hough transform search for
  continuous gravitational waves from isolated neutron stars assuming
  the input to be short segments of Fourier transformed data. The
  method presented here takes into account possible non-stationarities
  of the detector noise and the amplitude modulation due to the motion
  of the detector. These two effects are taken into account for the
  first stage only, i.e. the peak selection, to create the
  time-frequency map of our data, while the Hough transform itself is
  performed in the standard way.
\end{abstract}

\section{Introduction}
The Hough transform is a pattern recognition algorithm which was
originally invented to analyze bubble chamber pictures from CERN
\cite{hough1}. It was later patented by IBM \cite{hough2}, and it has
found many applications in the analysis of digital images \cite{ik}.
A detailed discussion of the Hough transform as applied to the search
for continuous gravitational waves can be found in \cite{hough04}.
An example of such a search is \cite{houghS2} in which this semi-coherent
technique is used to perform an all-sky search for isolated 
spinning neutron stars using two months of data collected in early 2003
from the second science run of the LIGO detectors \cite{ligo1, ligo2}.
The main results of \cite{houghS2} are all-sky upper limits on a set
of narrow frequency bands within the range $200$-$400\,$Hz and
including one spin-down parameter. The best upper limit on 
the gravitational wave  strain amplitude that we obtained in this 
frequency range is $4.43\times 10^{-23}$. Several searches have 
been completed or are underway using the same data. Examples of such
searches are \cite{cw-prl} which targets known radio pulsars at twice
the pulsar frequency, and \cite{fdsS2} that performs an all sky search
in a wide frequency band, but using only about 10 hours of data and applying
coherent (matched filtering) techniques.  Such techniques are very
computationally intensive for large parameter space searches.

The ultimate goal for wide parameter space searches for continuous 
signals over large data sets is to employ hierarchical schemes 
\cite{bc,pss01,cgk, f2000,fap,fp}
which alternate coherent and semi-coherent techniques. 
The Hough transform search would then be used to select candidates in 
parameter space to be followed up. It is crucial
that the first  stage of the search is as efficient as possible, 
since it determines the final sensitivity of the full pipeline.

In this paper we present an improved implementation of the Hough transform 
with respect to what was described in  \cite{hough04} when this is applied
to a set of Fourier transformed data (without any demodulations). This new
implementation takes into account non-stationarities of the noise floor
and also the amplitude modulation due to the antenna pattern that were
previously ignored.

The plan of the paper is as following: Section \ref{sec:waveform} described the 
waveforms we are looking for. Section \ref{sec:hough1} describes the basics of
the standard Hough transform and in Section \ref{sec:hough2} how this could be
improved in the presence of non-stationarities.

\section{The signal at the detector}
\label{sec:waveform}

The output of the detector can be represented by
\be
x(t) = h(t) + n(t) \, ,
\ee
where  $n(t)$ is the noise that
affects the observation at a detector time $t$, 
and $h(t)$ is the gravitational wave signal received at the detector:
\begin{equation}
h(t) = F_+({\bf n},\psi)h_+(t) + F_\times({\bf n},\psi)h_\times(t) \, .
\end{equation}
Here  $h_+$ and $h_\times$ are the two independent polarizations of the strain
amplitude at the detector  and $F_+$ and $F_\times$ are the antenna beam
pattern functions, that depend on the direction $\bf{n}$ to the star and also 
on the polarization angle $\psi$. Due to the motion of the Earth, the 
$F_{+,\times}$ depend implicitly on time. In case of a laser interferometer 
detector with perpendicular arms, the expressions for $F_+$ and
$F_\times$ are given by \cite{jks}: 
\be
F_+= a(t)\cos 2\psi+ b(t) \sin 2\psi \, , \quad
F_{\times}= b(t)\cos 2\psi - a(t) \sin 2\psi \, ,
\ee
where the functions $a(t)$ and $b(t)$ are independent of $\psi$.

We shall make the standard assumption that over a certain time-scale,
the noise $n(t)$ is stationary, although in realistic cases this
hypothesis is likely to be violated at some level. Within this
approximation, the Fourier components $\tilde{n}(f)$ of the noise 
are statistically described by:
\be
E[\tilde{n}(f)\tilde{n}^*(f')] =  \frac{1}{2} \delta (f-f')S_n(f) \, ,
\ee
where $E[]$ denotes the expectation value with respect to an ensemble of noise
realization, the $*$ superscript denotes complex conjugate,
$S_n(f)$ is the one sided  noise power spectral density, and
tildes denote Fourier transforms.

The form of the gravitational wave emitted by an isolated pulsar depends 
on the physical mechanism which may cause the neutron star to emit periodic
gravitational waves.
The main possibilities considered in the literature are
(i)~non-axisymmetric distortions of the solid part of the star,
(ii)~unstable $r$-modes in the fluid,
 and (iii)~free precession (or `wobble') 
(see \cite{S1-CW} and references therein).
If we assume the emission mechanism is (i), then the gravitational
waves are emitted at a frequency which is twice the rotational rate
$f_r$ of the pulsar.  Under this assumption, the waveforms for the two 
polarizations $h_{+,\times}$ are given by \cite{jks}:
\begin{equation}
h_+ = A_+\cos\Phi(t) = h_0 \frac{1+ \cos^2\iota}{2}\cos\Phi(t)\,, \qquad
h_\times = A_\times\sin\Phi(t) = h_0 \cos\iota\sin\Phi(t)\,,
\end{equation}
where $\iota$ is the angle between the pulsar's spin axis and the
direction of propagation of the waves, and $h_0$ is the amplitude:
\begin{equation} \label{eq:h0} h_0 = \frac{16\pi^2G}{c^4}\frac{I_{zz}\epsilon
f_r^2}{d}\,. \end{equation}
Here $G$ is Newton's gravitational constant, $c$ the speed of light, 
 $I_{zz}$ is the $z$-$z$ component of the star's moment of inertia with 
 the $z$-axis being its spin axis,
 $\epsilon:= (I_{xx}-I_{yy})/I_{zz}$ is the equatorial ellipticity of the
star, and $d$ is the distance of the star from Earth.

The phase $\Phi(t)$ takes its simplest form in the Solar System Barycenter
(SSB) frame where it can be expanded in a Taylor series:
\begin{equation}
 \label{eq:phasemodel}
\Phi(t) = \Phi_0 + 2\pi \sum_{n\ge 0} \frac{ f_{(n)}}{(n+1)!}(T - T_0)^{n+1} \,.
\end{equation}
Here $T$ is time in the SSB frame and $T_0$ is a fiducial start time.
The phase $\Phi_0$, frequency $f_{(0)}$ and the spin-down parameters $f_{(n)}$,
$n>0$,  
are defined at this fiducial start time. 
Neglecting relativistic effects which do not affect us significantly
in this case,
 the
relation between the time of arrival $T$ of the wave in the SSB frame
and in the detector frame $t$ is
\begin{equation}
T = t + \frac{\mathbf{n\cdot r}}{c}\,,
\end{equation}
where $\bf{n}$ is the direction to the pulsar and  $\bf{r}$ is the 
detector position in the SSB frame.
The instantaneous frequency $f(t)$ of the wave as
observed by the detector is given, to a very good approximation, by
the familiar non-relativistic Doppler formula:
\begin{equation}\label{eq:master}
f(t) - \hat{f}(t) = \hat{f}(t)\frac{ {\bf v} (t)\cdot\bf{n}}{c}
\end{equation}
where  ${\bf v}(t)$ is the
velocity of the detector at the detector  time $t$, 
 $\hat{f}(t)$ is the instantaneous signal frequency at
time $t$:
\begin{equation} 
\label{eq:fhat}
\hat{f}(t) = f_{(0)} + \sum_{n>0} \frac{ f_{(n)}}{n!}(T - T_0)^{n} \,.
\end{equation}
Equations (\ref{eq:master}) and (\ref{eq:fhat}) describe the
time-frequency pattern produced by a signal, and this is the pattern
that the Hough transform is used to look for.


\section{Standard Hough transform }
\label{sec:hough1}

The Hough transform can be used to find the pattern produced by the
Doppler shift (\ref{eq:master}) and the spin-down (\ref{eq:fhat}) of
a gravitational wave signal in the time-frequency plane of our data.
The parameters which determine this pattern are $(\{f_{(n)}\},\bf{n})$. 
The parameter space is covered by a discrete cubic grid and the result of the
Hough transform is an histogram for each point of this grid. 
By ``standard'' implementation of the Hough transform, we refer to the
implementation used in \cite{houghS2} where 
the starting point are $N$ short segments of Fourier transformed data
(which are called Short Fourier Transforms, or SFTs) with $\Tcoh$
being the time baseline of the SFTs. From this set of SFTs,
calculating the periodograms (the square modulus of the Fourier
transform) and selecting frequency bins (peaks) above a certain
threshold $\rho_\th$, we obtain a time-frequency map of our data.
Assuming a fixed threshold, the optimal value turns out to be
$\rho_\th=1.6$, corresponding to a peak selection probability of $q =
e^{-\rho_\th} = 0.2$ in the absence of a signal \cite{hough04}.
For each selected bin in the SFTs, we find which points in parameter
space are consistent with it, according to Eq.~(\ref{eq:master}), and
the number count in all such points is increased
by unity. This is repeated for all the selected  bins in all the
SFTs to obtain the final histogram.

There are several criteria to select the frequency bins. One of them
is to select by setting a threshold $\rho_\th$ on the
normalized power $\rho_k$ defined as
\begin{equation} \label{eq:normpower}
\rho_k = \frac{2|\tilde{x}_k|^2}{\Tcoh S_n(f_k)} \,,
\end{equation}
 where ${\tilde{x}_k}$ is the discrete Fourier transform of the data,
the frequency index $k$ corresponds to a physical frequency of $f_k= k/\Tcoh$,
and $S_n(f_k)$ is the single sided power spectral density of the
detector noise.
The $k^{th}$ frequency bin is selected if $\rho_k
\geq \rho_\th$, and rejected otherwise.  In this way, each SFT is replaced
by a collection of zeros and ones called a peak-gram. 
 The Hough transform is used to calculate the number count $n$ at each 
 parameter space point starting from this collection of peak-grams, i.e. each 
point in parameter space corresponds to a pattern in the time-frequency plane 
of our data, and the number count $n$  is the sum of the ones and zeros 
of the different peak-grams along this curve.

Let $p(n)$ be the
probability distribution of $n$ in the absence of a signal, and
$p(n|h)$ the distribution in the presence of a signal $h(t)$.  It is
clear that $0\leq n \leq N$, where $N$ is the number of SFTs, and it
can be shown that for stationary
Gaussian noise, $p(n)$ is a binomial distribution with mean
$Nq$ where $q = e^{-\rho_\th}$ is the probability that any frequency
bin is selected:
\begin{equation}
\label{eq:binomialnosig}
p(n) = \left( \begin{array}{c} N \\ n \end{array}  \right)
q^n(1-q)^{N-n}\,.
\end{equation}
In the presence of a signal, the distribution $p(n|h)$ is
ideally also a binomial but with a slightly larger mean $N\eta$
where, for weak signals, $\eta$ is given by
\begin{equation}
\label{eq:eta}
\eta = q\left\{1+\frac{\rho_\th}{2}\lambda_k +
\mathcal{O}(\lambda_k^2)  \right\}\,.
\end{equation}
$\lambda_k$ is the signal to noise ratio within a single SFT, and
for the case when there is no mismatch between the signal and the
template:
\begin{equation}
\label{eq:lambda}
\lambda_k = \frac{4|\tilde{h}(f_k)|^2}{\Tcoh S_n(f_k)}
\end{equation}
with $\tilde{h}(f)$ being the Fourier transform of the signal $h(t)$
(see \cite{hough04} for details)

%
Candidates in parameter space are selected by setting a threshold
$n_\th$ on the number count. For the case of large $N$, and Gaussian stationary
noise, the relation between the the false alarm rate $\alpha$ and the threshold
$n_\th$ is independent of the signal strength and given by:
\begin{equation}
n_\th= Nq +
\sqrt{2N q(1-q)}\,\textrm{erfc}^{-1}(2\alpha)
\end{equation}
where,  $\textrm{erfc}^{-1}$ is the inverse of the complementary error
function. On the average, the  weakest signal which will cross the above
thresholds at a false alarm rate $\alpha$ and false dismissal
$\beta$ is given by
\begin{equation} 
\label{eq:sensitivity}
h_0 = 5.34 \frac{\mathcal{S}^{1/2}}{N^{1/4}}\sqrt\frac{S_n}{\Tcoh}\,, \quad
\mathcal{S} = \textrm{erfc}^{-1}(2\alpha) + \textrm{erfc}^{-1}(2\beta)\,.
\end{equation}
Equation (\ref{eq:sensitivity}) gives the smallest signal which can be
detected by the search.

\section{Improved peak selection for the Hough transform }
\label{sec:hough2}

The approximation that the distribution $p(n|h)$ in the
presence of a signal is binomial can break down for several reasons:
the random mismatch between the signal and the template used to calculate the
number count, non-stationarity of the noise, and the amplitude modulation 
of the signal which causes $\lambda_k$  to vary from one SFT to another and 
for different sky locations and pulsar orientations. The result of these 
effects is to ``smear'' out the binomial distribution in the presence of 
a signal as it is illustrated in Figure~14 in \cite{houghS2}.
%
%
To account for these non-stationarities a possible {\it Adaptive Hough
  Transform} is discussed in \cite{paf} in which Palomba \etal suggest
not to increase the number-count by unity, but by a real number
depending on the detector response function and the noise level.

In this paper we propose an alternative method that consists in
setting an adaptive peak selection threshold 
that varies for the different SFTs and sky locations, and then
performing the {\it Standard Hough Transform} on the resulting
peak-grams.  We expect this method should have a somewhat similar gain
in sensitivity as the method presented in \cite{paf}, but with a
better performance since all the implementation tricks described in
\cite{hough04} and \cite{houghS2} Sec~VI.C can still be applied.
However, we emphasize that a proper statistical justification of this
method is still unclear and is being investigated.  

\subsection{Dealing with non-stationarity}

Let us consider the quantity $\rho_k\lambda_k /2$,  related
to the second summand in right hand side of Eq.~(\ref{eq:eta}), which, 
in the discrete case, 
can be rewritten as:
\be
\label{eq:lrho}
\frac{\rho_k \lambda_k}{2} = \frac{|\tilde{x}_k|^2}{ E[ |\tilde{n}_k|^2]
} \, \frac{|\tilde{h}(f_k)|^2}{ E[ |\tilde{n}_k|^2]} =
\frac{|\tilde{x}_k|^2}{ E[ |\tilde{n}_k|^2] } \, \frac{\langle E[
  |\tilde{n}_k|^2] \rangle_\N}{ E[ |\tilde{n}_k|^2]} \,
\frac{|\tilde{h}(f_k)|^2}{\langle E[ |\tilde{n}_k|^2]\rangle_\N} \, .
\ee Here $E[ |\tilde{n}_k|^2] = \Tcoh S_n(f_k) /2$ denotes the noise
contribution in a single SFT, which is assumed to be stationary in a
time scale $\Tcoh$ although it can vary from one SFT to another.  In
practice the expected noise level, in each data stretch, is estimated
from the data itself by means of the running median \cite{mohanty02b}
and it is frequency dependent.  $\langle E[|\tilde{n}_k|^2]\rangle_\N$
denotes the average noise floor of the $N$ different SFTs.

We typically choose $\Tcoh = 30\,$min.  While a larger value of
$\Tcoh$ would lead to better sensitivity, this time baseline cannot be
made arbitrarily large because of the frequency drift caused by the
Doppler effect and the spin-down; we would like the signal power of a
putative signal to be concentrated in less than half the frequency
resolution $1/\Tcoh$.  It is shown in \cite{hough04} that at
$300\,$Hz, we could ideally choose $\Tcoh$ up to $\sim60\,$min.  On
the other hand, we should be able to find a significant number of such
data stretches during which the interferometers are in lock, the noise
is stationary, and the data are labeled satisfactory according to
certain data quality requirements.

In order to keep $\rho_\th \lambda_k / 2 $ constant across the different
$N$ data segments, Eq.~(\ref{eq:lrho}) suggests that, ignoring the
amplitude modulation, a more natural threshold for the peak selection
should be placed not on $\rho_k$ but on 
\begin{equation} \hat \rho_k=
\frac{|\tilde{x}_k|^2}{ E[ |\tilde{n}_k|^2] } \, \frac{\langle E[
  |\tilde{n}_k|^2] \rangle_\N}{ E[ |\tilde{n}_k|^2]} \, .  
\end{equation} 
The probability of selecting a peak in the absence of a signal in a
single SFT would then be:
\begin{equation} \hat q_k= \exp \left[ - \rho_\th \, \frac{ E[
    |\tilde{n}_k|^2]} {\langle E[ |\tilde{n}_k|^2] \rangle_\N} \right]
\end{equation} 
thus selecting fewer peaks in the most noisy periods.

The presence of very noisy periods, if those are not vetoed for the
analysis, could in principle disturb not just some SFTs but also
dominate the average $\langle E[ |\tilde{n}_k|^2]\rangle_\N$, and this
could degrade the sensitivity. Therefore, in practice, $\langle
E[|\tilde{n}_k|^2]\rangle_\N$ should be estimated by a more
sophisticated method, e.g. based on the mean of some sorted points
between different quartiles of the different SFTs, or the harmonic
mean etc.

\subsection{Amplitude modulation corrections}

In Eq.~(\ref{eq:lambda}) the amplitude of the signal
$|\tilde{h}(f_k)|^2$ is proportional to $F_+^2A_+^2 +
F_\times^2A_\times^2$, again taking $a$ and $b$ to be approximately
constant.  Averaging over the polarization angle $\psi$ \be \langle
\cdots\rangle_\psi :=\frac{1}{2\pi} \int_0^{2\pi} d\psi (\cdots) \, ,
\ee we have $\langle F_+^2\rangle_\psi = \langle
F_\times^2\rangle_\psi = (a^2+b^2)/2$.  Therefore
$\langle|\tilde{h}(f_k)|^2\rangle_\psi \propto a^2+b^2$ and this holds
true independently of the polarization of the waveform, and also
$|\tilde{h}(f_k)|^2 \propto a^2+b^2$ for circular polarization ($\cos
\iota =1$) independently of $\psi$.

The functions $a$ and $b$ vary in time due to the motion of the
detector, although they can be considered constant for the typical
time baseline $\Tcoh$ of the SFTs.  This suggests that the peak
selection criteria should be based on setting a threshold on:
\begin{equation}
  \bar \rho_k= \frac{|\tilde{x}_k|^2}{ E[
    |\tilde{n}_k|^2] } \, \frac{\langle E[ |\tilde{n}_k|^2] \rangle_\N}{
    E[ |\tilde{n}_k|^2]} \, \frac{a^2+b^2}{\langle a^2+b^2 \rangle_\N}
  \, , 
\end{equation}
and selecting those frequency bins (peaks) which satisfy $\bar \rho_k
> \rho_\th$. The peak selection probability in the absence of a signal
would be: \be \bar q_k= \exp \left[ - \rho_\th \, \frac{ E[
    |\tilde{n}_k|^2]} {\langle E[ |\tilde{n}_k|^2] \rangle_\N} \,
  \frac{\langle a^2+b^2 \rangle_\N}{a^2+b^2} \right] \, .  \ee In
other words, the threshold on ${|\tilde{x}_k|^2}/{ E[ |\tilde{n}_k|^2]
} $ would be frequency dependent and also different for different SFTs
and sky locations.

Since the functions $a$ and $b$ vary for different sky locations and
we want to take advantage of the implementation method described in
\cite{hough04}, for a wide area or an all-sky search, the celestial
sphere should then be divided into several smaller regions.  For each
of these small sky-patches, the functions $a$ and $b$ should be
evaluated at the center of the sky-patch and assumed constant for the
entire sky-patch.  The peak selection procedure needs to be repeated
for each individual sky-patch we want to analyze.  The smaller the
patch the more the gain in sensitivity and also the smaller the
distortions induced by the tiling of the celestial as discussed in
\cite{houghS2}.
 
The details of the statistical properties of this search will be
presented elsewhere.  In particular, as mentioned earlier, a proper
statistical justification of this method in terms of, say, a
Neyman-Pearson criteria, the regime where the gain in sensitivity is
most significant, and a large scale Monte-Carlo study are currently
being investigated.

\ack We would like to thank the LIGO Scientific Collaboration for many
useful discussions.  We also acknowledge the support of the Max-Planck
Society and the Spanish {\it Ministerio de Educaci\'on y Ciencia}
Research Project REF: FPA-2004-03666.  AMS is grateful to the Albert
Einstein Institute for hospitality where this work was initiated.

\end{document}